\documentclass[aps,prb,
 amsmath,amssymb,twocolumn, groupedaddress]{revtex4-1}
\usepackage{blindtext}
\usepackage{natbib}
\usepackage[dvipdfmx]{graphicx}

\usepackage{color}
\usepackage{comment}

\begin{document}

\title{Electronic states of multilayer VTe$_2$: quasi-one-dimensional Fermi surface and implications to charge-density waves}
\author{Tappei Kawakami,$^1$ Katsuaki Sugawara,$^{1, 2, 3, 4}$ Takemi Kato,$^1$ Taiki Taguchi,$^1$ Seigo Souma,$^{2,3}$ Takahashi Takashi,$^{1, 2, 3}$ and Takafumi Sato,$^{1, 2, 3, 5}$}
\affiliation{$^1$Depertment of Physics, Tohoku University, Sendai 980-8578, Japan}
\affiliation{$^2$Center for Spintronics Reaserch Network, Tohoku University, Sendai 980-8578, Japan}
\affiliation{$^3$WPI Reaserch Center, Advanced Institute for Materials Reaserch, Tohoku University, Sendai 980-8577, Japan}
\affiliation{$^4$Precursory Research for Embryonic Science and Technology (PRESTO), Japan Science and Technology Agency (JST), Tokyo 102-0076, Japan}
\affiliation{$^5$International Center for Synchrotron Radiation Innovation Smart (SRIS),
Tohoku University, Sendai 980-8577, Japan}
\date{\today}

\begin{abstract}
We have performed angle-resolved photoemission spectroscopy on epitaxial VTe$_2$ films to elucidate the relationship between the fermiology and charge-density waves (CDW). We found that a two-dimensional triangular pocket in 1 monolayer (ML) VTe$_2$ is converted to a strongly warped quasi-one-dimensional (1D) Fermi surface in the 6ML counterpart, likely associated with the 1$T$-to-1$T$' structural phase transition. We also revealed a metallic Fermi edge on the entire Fermi surface in 6ML at low temperature distinct from anisotropic pseudogap in 1ML, signifying a contrast behavior of CDW that is also supported by first-principles band-structure caluculations. The present result points to the importance of simultaneously controlling the structural phase and fermiology to manipulate the CDW properties in ultrathin transition-metal dichalcogenides.
\end{abstract}

\maketitle


\section{INTRODUCITON}\label{INTRODUCITON}
   The discovery of Dirac electrons in graphene has initiated the search for new types of two-dimensional (2D) materials which exhibit quantum properties distinct from those of bulk counterparts \cite{NovoNature2005, Mak3PRL2010, XiaoPRL2012, CaoNatureCom2012, ZengNatureNanotech2012, Mak1NatureNanotech2012, WangPRL2012, Mak2Science2014, XiNaturePhysics2016, NakataNPGAsiaMaterials2016, NakataACSApplNanoMater2018}. Atomically-thin transition-metal dichalcogenides (TMDs) are attracting particular attention not only because parent bulk TMDs exhibit various physical properties, but also because the two-dimensionalization of TMDs leads to outstanding characteristics such as Ising superconductivity \cite{XiNaturePhysics2016}, spin-valley Hall effect \cite{Mak1NatureNanotech2012}, and 2D Mott-insulating phase \cite{NakataNPGAsiaMaterials2016, NakataACSApplNanoMater2018}. To fully explore unique features of 2D TMDs, it is essential to compare the electronic properties between 2D and 3D systems and obtain insight into the evolution of physical properties upon the dimensional crossover. In this context, one may expect a similar electronic property between 2D and 3D cases because bulk TMDs are a layered quasi-2D system. In fact, similarity between monolayer and bulk has been identified in some TMDs, as highlighted by the 2$\times$2 charge-density wave (CDW) in 1$T$-TiSe$_2$ accompanied by the exciton condensation \cite{ChenNatureCommun2015, SugawaraACSNano2016}, the superconductivity coexisting with CDW in 1$H$/2$H$-NbSe$_2$,\cite{Mak1NatureNanotech2012} and the CDW associated with Star-of-David clusters in 1$T$-TaSe$_2$.\cite{ChenNatreuPhysics2020} On the other hand, some 2D TMDs exhibit electronic properties that cannot be understood simply by extending the framework of 3D system, as represented by monolayer (ML) 1$T$-VSe$_2$ with a variety of periodic lattice distortions (PLD; e.g. 4$\times$4, 4$\times$1, $\sqrt{7}{\times}\sqrt{3}$, 4$\times\sqrt{3}$) \cite{ZhangPRMat2017, FengNanoLett2018, DuvjirNanoLett2018, UmemotoNanoRes2019, ChenPRL2018} which are hardly explained in terms of a simple Fermi-surface (FS) nesting picture, different from the nesting-driven 4$\times4{\times}{\sim}3$ CDW in bulk 1$T$-VSe$_2$. \cite{TsutsumiPRB1982, TerashimaPRB2003, SatoJPhysSocJpn2004, StrocovPRL2012, Pasztor2DMater2017} 

VTe$_2$ is an excellent target to address the essential question regarding the evolution of physical properties upon the 2D-3D crossover, because the electronic property shows an intriguing difference between monolayer and bulk. Bulk VTe$_2$ forms the monoclinic 1$T$'' phase with 3$\times$1$\times$3 PLD and undergoes a structural transition to the polymorphic 1$T$ structure at $T$ = 482 K \cite{OhtaniSolidStateCommun1981, BronsemaJSolidStateChem1984, MitsuishiNatureCommun2020}. On the other hand, monolayer VTe$_2$ stabilizes with the 1$T$ structure \cite{SugawaraPRBR2019, WongACSNano2019, WangPRBR2019, MiaoPRB2020, LiuNanoResearch2019, WuPRB2020} and exhibits the 4$\times$4 CDW \cite{OhtaniSolidStateCommun1981, BronsemaJSolidStateChem1984, MitsuishiNatureCommun2020}. Although the relationship between fermiology and PLD as well as the origin of different crystal structures between monolayer and bulk have been intensively discussed, these key questions \cite{CoelhoJPhysChemLett2019, LasekACSNano2020} are still controversial \cite{CalandraPRB2009, DaiJphysChem2019, CoelhoJPhysChemLett2019, LasekACSNano2020}. Previous studies also reported that multilayer VTe$_2$ exhibits the 2$\times$1 periodicity distinct from bulk and monolayer\cite{CoelhoJPhysChemLett2019, LasekACSNano2020, DaiJphysChem2019}, but the consensus on the origin of the 2$\times$1 periodicity and its link to possible CDW properties has not yet been obtained. Dai $et$ $al$. successfully fabricated multilayer VTe$_2$ films by the molecular beam epitaxy (MBE) and suggested the multiple CDW phase that depends on the number of VTe$_2$ layers \cite{DaiJphysChem2019}, whereas Coelho $et$ $al$. did not observe CDW in the films with the thickness of more than bilayer. Coelho $et$ $al$. suggested the 1$T$' structure as an origin of the 2$\times$1 periodicity and associated its stabilization with the charge transfer from Te to V due to the interlayer interaction\cite{CoelhoJPhysChemLett2019}. On the other hand, Lasek $et$ $al$. suggested the formation of V$_3$Te$_4$ phase with the 2$\times$1 periodicity, based on their observation that MBE-grown multilayer TMD films suffer spontaneous intercalation of transition-metal atoms\cite{LasekACSNano2020}. These experiments highlight that, although the relationship between fermiology and PLD as well as the change in the crystal structure from monolayer to bulk has been intensively discussed, the key question regarding the relationship between the number of layers and electronic properties is still controversial\cite{CoelhoJPhysChemLett2019, LasekACSNano2020, DaiJphysChem2019}. In a broader perspective, a general guideline to connect physics of 2D and 3D TMDs is still missing. These issues necessitate an investigation of electronic states in $multilayer$ VTe$_2$.

In this article, we report angle-resolved photoemission spectroscopy (ARPES) study of monolayer and multilayer VTe$_2$ films grown epitaxially on bilayer graphene, combined with the first-principles band-structure calculations. We found that the FS of 6ML film is composed of a poorly nested quasi-1D FS with a metallic Fermi edge, drastically different from that of 1ML film characterized by a well-nested triangular pocket. A comparison of the band structure between ARPES and band calculations supports the 1$T$' structure for 6ML film, distinct from the 1$T$'' (1$T$) structure of bulk (monolayer). The present result demonstrates a crucial role of interlayer effect to manipulate the crystal phase and fermiology in ultrathin TMDs.

\section{EXPERIMENTS}
Monolayer and multilayer VTe$_2$ films were grown by the MBE method. ARPES measurements were carried out at Photon Factory BL28 of KEK. First-principles band-structure calculations were carried out by using the Quantum Espresso code package \cite{GiannozziJPhysCondesMatter2009} with generalized gradient approximation \cite{PerdewPRL1996}. The calculation of FS was performed by using Wannier90 code \cite{PizziJPhysCondes2020}. For details, see Appendix \ref{App. A}.

\section{RESULTS AND DISCUSSION}
\subsection{Characterization of VTe$_2$ films}
First, we show characterization of VTe$_2$ films. Figures 1(a) and 1(b) show the low-energy-electron-diffraction (LEED) pattern of 1 and 6ML VTe$_2$ films, respectively, grown on bilayer graphene/4$H$-SiC(0001). We clearly recognize in Fig. 1(a) a sharp 1$\times$1 spot associated with the formation of 1ML VTe$_2$ film \cite{SugawaraPRBR2019, NakataNPGAsiaMaterials2016, NakataACSApplNanoMater2018, UmemotoNanoRes2019}, together with the graphene spots. On the other hand, when the film thickness is increased to 6ML [Fig. 1(b)], the 2$\times$2 spot (red arrow) emerges and the graphene spot disappears. This cannot be explained in terms of the formation of monoclinic (1$T$'') phase seen in bulk VTe$_2$ \cite{OhtaniSolidStateCommun1981, BronsemaJSolidStateChem1984}, because the periodicity of 1$T$'' phase is 3$\times$1.


\begin{figure}[htbp]
\begin{center}
\includegraphics[width=2.9in]{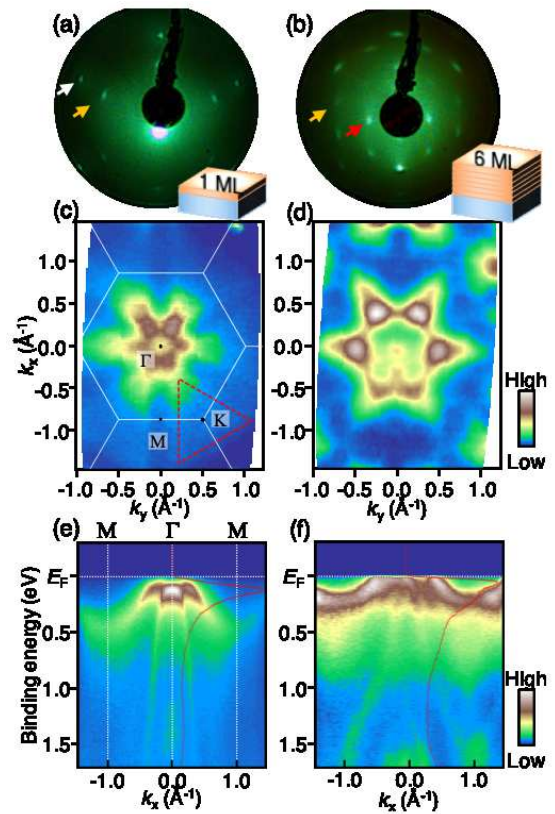}
\caption{(a), (b) LEED patterns of 1 and 6ML VTe$_2$, respectively. Orange, red, and white arrows represent the spots/streaks from 1$\times$1 VTe$_2$, 2$\times$1 VTe$_2$, and bilayer graphene, respectively. (c), (d) ARPES-intensity plots at $E_{\rm{F}}$ of 1 and 6ML VTe$_2$, respectively, measured at $T$ = 40 K as a function of 2D wave vectors. $k_x$ and $k_y$ are defined according to the BZ of BL-graphene substrate. White solid lines in (c) represent the BZ of 1$T$ phase. Dashed red triangle highlights the location of triangular FS. (e), (f) ARPES-intensity plots of 1 and 6ML VTe$_2$ along the $k_x$ cut measured with $h\nu$ = 74 eV at $T$ = 40 K. EDC at $\Gamma$ is also shown by red curve.}
\label{fig:fig1}
\end{center}
\end{figure}


Next we examine the electronic states by ARPES. Figures 1(c) and 1(d) display the plot of ARPES intensity at $E_{\rm{F}}$ for 1 and 6ML, respectively, measured at $T$ = 40 K. In 1ML, one can recognize a large triangular FS centered at the $\rm{K}$ point in the Brillouin zone (BZ) of 1$T$ phase, with an enhanced intensity around the corner of triangular pocket near the $\Gamma$ point \cite{SugawaraPRBR2019, WangPRBR2019}. On the other hand, the ARPES intensity for 6ML displays a snow-flake shape centered at $\Gamma$. Also, the intensity profile away from $\Gamma$ (e.g. $|k_x|$ $>$ 1.0 $\rm{\AA}^{-1}$, $|k_y|$ $\sim$ 1.0 $\rm{\AA}^{-1}$) is very different between 1 and 6ML. Such a difference in the FS topology is also reflected in the experimental band dispersion shown in Figs. 1(e) and 1(f). Namely, the bottom of V 3$d$ band is located at the binding energy ($E_{\rm{B}}$) of $\sim$0.5 eV at the M point for 1ML, while it is pushed upward by $\sim$0.3 eV for 6ML, resulting in the narrower band width along the $k_x$ cut. Moreover, although the V 3$d$ band for 1ML does not touch $E_{\rm{F}}$ along the $k_x$ cut [Fig. 1(e)], it apparently crosses $E_{\rm{F}}$ for 6ML [Fig. 1(f)]. These results indicate that the electronic structure strongly depends on the number of VTe$_2$ layer.

\subsection{Fermiology and band structure}
We compare in Figs. 2(a)-2(d) the experimental FS mapping and the band dispersion measured along the $k_y$ cut for 6ML [Figs. 2(a) and 2(b)], compared with the band-structure calculations for 1ML VTe$_2$ by assuming the 1$T$ structure [Figs. 2(c) and 2(d)] (see Fig. 5 in Appendix \ref{App. B}). One can recognize from a comparison of Figs. 2(a) and 2(c) that the calculated large triangular FS centered at $\Gamma$ has no counterpart in the experiment. Such a disagreement is also visible in the band dispersion; as shown in Figs. 2(b) and 2(d), although one can commonly recognize in the experiment and calculation the Te 5$p$ bands with a holelike dispersion centered at $\Gamma$, a band showing a shallower energy dispersion below $E_{\rm{F}}$ in the experiment is not seen in the calculation. This is because the calculated V 3$d$ band is pushed upward into the unoccupied region along this $\bf{k}$ cut [Fig. 2(d)]. These results strongly suggest that the 6ML film does not take the 1$T$ structure unlike the case of 1ML, supporting the previous studies\cite{CoelhoJPhysChemLett2019, LasekACSNano2020, DaiJphysChem2019}.

Next, we discuss a possible cause of the lattice-periodicity doubling. One may think that V atoms are intercalated within the van der Waals gap and show the 2$\times$2 order \cite{NakanoNanoLett2019}. Such intercalation would cause heavy electron doping to VTe$_2$ layers. However, the observed band dispersion are found to be incompatible with the calculations for the electron-doped 1$T$ phase. Moreover, our transmission-electron-microscopy (TEM) measurements revealed no clear signature of periodically ordered intercalated V atoms in the van der Waals gap (see Fig. 6 in Appendix \ref{App. C}). As a cause for the 2$\times$2 periodicity, one can consider the 1$T$' structure that possess a 2$\times$1 periodicity in a single domain \cite{CoelhoJPhysChemLett2019}. It is expected from the symmetry mismatch between 1$T$'-VTe$_2$ ($\rm{C_2}$) and bilayer graphene ($\rm{C_6}$) that three types of 1$T$' domains rotated by 120${}^\circ$ with each other are formed on bilayer graphene. This would lead to observation of the 2$\times$2 periodicity in the LEED pattern. As shown in Fig. 2(e), the calculated FS for the single-domain monolayer 1$T$'-VTe$_2$ consists of two quasi-1D open FS and two smaller pockets centered at the $\Gamma$ and $\rm{Y}$ points of 1$T$' BZ, respectively (it is noted that we have adopted the calculation for monolayer 1$T$'-VTe$_2$ because that for 6ML was difficult due to the large number of atoms involved in a unit cell; even when the calculation could be possible, the obtained band structure would be too complicated to be compared with the experimental data). When three domains are taken into account, the FS topology becomes complicated as shown in Fig. 2(g). This is also the case for the band dispersion along the $k_y$ cut in graphene BZ [Figs. 2(f) and 2(h)]. Although it is difficult to absolutely judge the amount of agreement between the experiment and calculation, the observed snow-flake-shaped intensity in Fig. 2(a) appears to trace the $\bf{k}$ region where the calculated FS contour has a high density in Fig. 2(g). The better agreement of experiments with the calculation for the 1$T$' phase is also collaborated by the energy-contour plot at several binding-energy slices (see Fig. 7 in Appendix \ref{App. D}). Also, the observed shallow V 3$d$ band below ${E_{\rm{F}}}$ in Fig. 2(b) which has no counterpart in the calculation for the 1$T$ phase [Fig. 2(d)] appears to exist in the calculation for the 1$T$' phase [Fig. 2(f)] although the bandwidth is narrower in the experiment probably due to the electron-correlation-driven band-renormalization effect \cite{SugawaraPRBR2019}. It is noted that the calculated small hole pocket at $\Gamma$ originating from the Te 5$p$ band is not clearly seen in the experiment because this band does not cross $E_{\rm{F}}$.

\onecolumngrid
\begin{figure*}
\begin{center}
\includegraphics[width=6.0in]{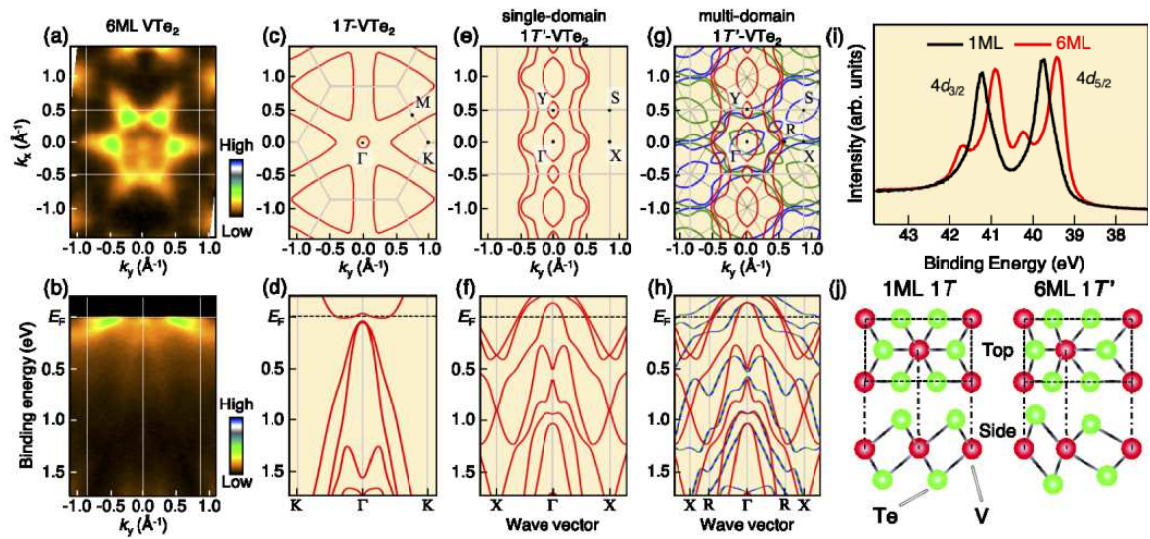}
\caption{(a), (b) FS mapping and ARPES-intensity plot along the $k_y$ cut for 6ML VTe$_2$, respectively. White solid lines represent the BZ boundaries of the single-domain 1$T$' phase. (c), (d) Calculated FS and band structure along the $\Gamma \rm{K}$ cut, respectively, obtained with band-structure calculations for 1ML 1$T$-VTe$_2$. (e), (f) Same as (c) and (d), but for single-domain 1ML 1$T$'-VTe$_2$. (g), (h) Same as (c) and (d), but for multi-domain 1ML 1$T$'-VTe$_2$. For 1$T$'-VTe$_2$, we defined the intersection of BZs from different domains as ``$\rm{R}$'' point. (i) EDC at $T$ = 40 K around the Te-4$d$ core levels measured with $h\nu$ = 74 eV for 1 (black curve) and 6ML (red curve) VTe$_2$. (j) Schematics of crystal structure for 1$T$ (left) and 1$T$' (right) VTe$_2$.}
\label{fig:fig2}
\end{center}
\end{figure*}
\twocolumngrid
 
The 1$T$' structure for 6ML is further corroborated by the core-level data. As shown in Fig. 2(i), the Te-4$d$ core-level spectrum for 1 ML consists of two peaks which correspond to the spin-orbit satellites of Te 4$d_{3/2}$ and $4d_{5/2}$ orbitals. On the other hand, each peak further splits into two peaks in 6ML. This is due to the existence of two kinds of chemical bond lengths surrounding the Te atom as naturally inferred from the 1$T$' structure, in contrast to the 1$T$ structure with a single bond length [Fig. 2(j)]. These results suggest that 6ML VTe$_2$ takes the 1$T$' structure, in line with Raman spectroscopy \cite{DaiJphysChem2019} and scanning tunneling microscopy \cite{CoelhoJPhysChemLett2019}. It is known that bulk VTe$_2$ stabilizes into the 1$T$'' phase with the 3$\times$1 periodicity \cite{OhtaniSolidStateCommun1981, BronsemaJSolidStateChem1984}, whereas our multilayer VTe$_2$ film takes the 1$T$' structure with the 2$\times$1 periodicity. One can thus expect that the 1$T$'' phase cannot be achieved by simply increasing the number of layers in the epitaxial film.

To clarify the FS topology of 6ML in more detail, we show in Figs. 3(b)-3(e) the ARPES intensity at $T$ = 40 K measured along several cuts (cuts A-D) in Fig. 3(a). One can see a shallow V 3$d$ band showing a holelike dispersion centered at $k_y$ = 0 which crosses $E_{\rm{F}}$ in all the $\bf{k}$ cuts (white arrows), suggesting the existence of an open FS. To examine this point in more detail, we show in Fig. 3(f) the momentum-distribution curves (MDCs) at $E_{\rm{F}}$ obtained along several $\bf{k}$ cuts shown by white dotted lines in Fig. 3(a). One can recognize that the MDCs along cuts 5-8 contain tiny peaks (marked by blue dots) which are smoothly connected to the prominent peaks along cuts 1-4 and 9-11, despite marked intensity suppression. As shown in Fig. 3(a), the FS estimated by tracing MDC peak positions in Fig. 3(f) shows an open quasi-1D character with a strong warping effect as opposed to 1ML with a simple 2D triangular pocket [Fig. 1(c)]. Besides the quasi-1D band, the ARPES intensity in Figs. 3(b)-3(e) shows additional bands (small gray arrows) which likely originate from the other two crystal domains rotated by $\pm$120${}^\circ$; we will not get into further detail because they are not the focus of the present study.

\onecolumngrid
\begin{figure*}
\begin{center}
\includegraphics[width=6in]{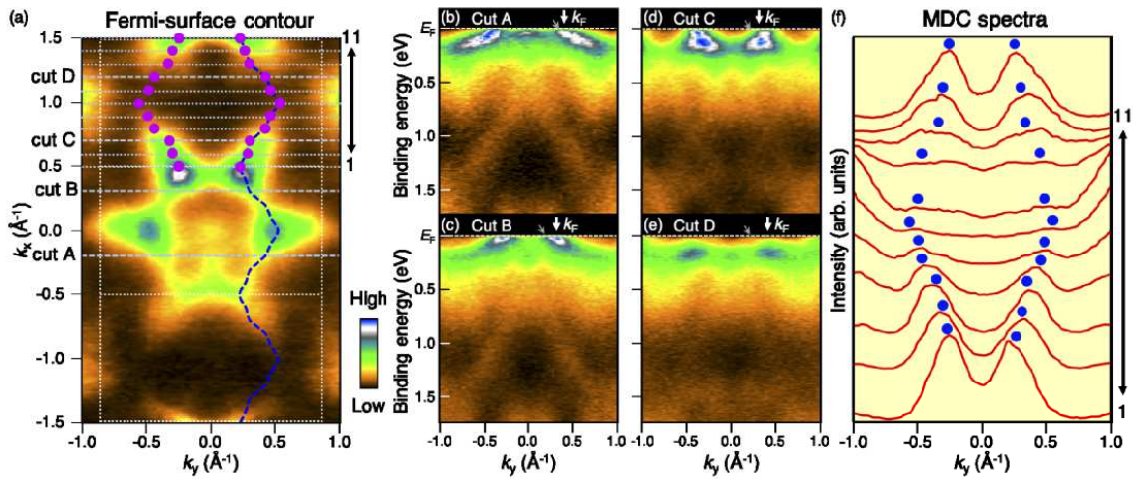}
\caption{(a) FS mapping of 6ML 1$T$'-VTe$_2$. Experimental Fermi wave vectors ($k_{\rm{F}}$'s; purple dots) estimated from the MDCs in (f) are overlaid by dashed blue curve. $k_{\rm{F}}$'s were determined by tracing the $\bm{k}$ location of $E_{\rm{F}}$-crossing points of energy bands along several $k_y$ slices with different $k_x$'s. (b)-(e) ARPES intensity measured along four representative $k$ cuts in (a) (cuts A-D). White arrows indicate the $k_{\rm{F}}$ point from the dominant crystal domain, whereas gray arrows represent possible contributions from the other two domains. (f) MDCs at $E_{\rm{F}}$ for 6 ML film plotted along several $\bf{k}$ cuts shown by white dotted lines in (a).}
\label{fig:fig3}
\end{center}
\end{figure*}
\twocolumngrid

Here, we comment on the observation of three types of 1$T$' domains in terms of the symmetry of ARPES intensity and spot size of incident photons. Since the beam spot size (200$\times$300 $\mu{m}^2$) is much larger than the crystal domain size (a few 10 $\mu{m}$) \cite{CoelhoJPhysChemLett2019}, we simultaneously observe electronic structures for three types of domains with nearly equal spectral weight. This is also consistent with our LEED measurements with the beam spot of $\sim$ 100$\times$100 $\mu{m}^2$ which signified uniform 2$\times$2 spots like Fig. 1(b) independent of the position of surface, without any indication of the local 2$\times$1 pattern. The mixture of three domains is also supported by the experimental fact that (i) the ARPES intensity distribution in our multilayer films always shows a similar intensity distribution among different samples and (ii) scanning of the beam position on the surface did not cause a meaningful change in the ARPES intensity. This suggests that the two-fold ARPES intensity pattern in Fig. 3(a) is due to the matrix-element effect of photoelectron intensity and does not reflect the genuine intensity distribution of the single-domain 1$T$' film. This argument is corroborated by the ARPES intensity for the monolayer 1$T$ sample in Fig. 1(c) which also shows the two-fold pattern, despite the six-fold symmetry of the 1$T$ structure. It is noted that a domain-selective ARPES experiment that utilizes a small beam spot focused down to a few $\mu{m}$ (micro-ARPES) would be useful to firmly establish the quasi-1D nature of the electronic states.

\subsection{Energy gap and CDW}

Now that the FS topology for 6ML is established, a next issue is whether or not an energy gap associated with CDW is realized. This point is crucial since previous studies on 1ML 1$T$-VTe$_2$ (and also 1$T$-VSe$_2$) reported a gap opening at low temperatures \cite{UmemotoNanoRes2019, SugawaraPRBR2019, WangPRBR2019}. We show in Figs. 4(a) and 4(b) the energy distribution curve (EDC) at $T$ = 40 K for 6ML measured at various $k_{\rm{F}}$ points on the quasi-1D FS. In the first BZ of single-domain 1$T$' phase [points 1-11; Fig. 4(a)], one can see a peak around $E_{\rm{F}}$ followed by a sharp Fermi-edge cut-off. The Fermi edge is also recognized in the second BZ [points 12-15; Fig. 4(b)] whereas the EDC is complicated by the presence of another peak at $E_{\rm{B}}$ $\sim$0.1-0.2 eV. These results suggest the absence of an energy gap on the entire quasi-1D FS at $T$ = 40 K, in contrast to the 1ML counterpart showing an anisotropic pseudogap \cite{SugawaraPRBR2019} highlighted in Fig. 4(c)
(for details, see Fig. 8 of Appendix \ref{App. E}). This pseudogap was initially explained in terms of CDW fluctuations in the previous ARPES study \cite{SugawaraPRBR2019}, whereas the subsequent STM and LEED experiments \cite{WangPRBR2019, MiaoPRB2020} reported the 4$\times$4 CDW and attributed the pseudogap to the CDW gap due to the FS nesting in the flat segments of triangular FS \cite{WangPRBR2019} (see also Appendix \ref{App. E}). We found from the analysis of reflection-high-energy-electron-diffraction (RHEED) patterns that the 1$T$' structure is immediately formed when the film thickness reaches 2ML during epitaxy (see Fig. 9 in Appendix \ref{App. F}). The observed intriguing difference in the gap-opening behavior between single- and multi-layer VTe$_2$ is explained in terms of the difference in their crystal phases (1$T$ vs 1$T$') and the change in FS topology and associated CDW properties. To further examine this possibility, we have calculated $\bf{q}$ (momentum transfer) dependence of the imaginary part of electronic susceptibility $\chi_0(\bf{q})$ \cite{JohannesPRB2008} based on the band calculations for 1ML 1$T$'-VTe$_2$, as shown in Fig. 4(d). One can see in Fig. 4(d) the absence of any sharp features except for $\bf{q}$ $\sim$ 0. This is reasonable because the calculated FSs do not obviously contain well-nested flat segments because of the warping effect, as also revealed in the ARPES experiment. Such behavior is in contrast to the 1$T$ case [Fig. 4(e)] where there exist distinct bright spots away from $\bf{q}$ $\sim$ 0 due to the better FS-nesting condition associated with the flat segments of triangular FS.

\begin{figure}[htbp]
\begin{center}
\includegraphics[width=3.4in]{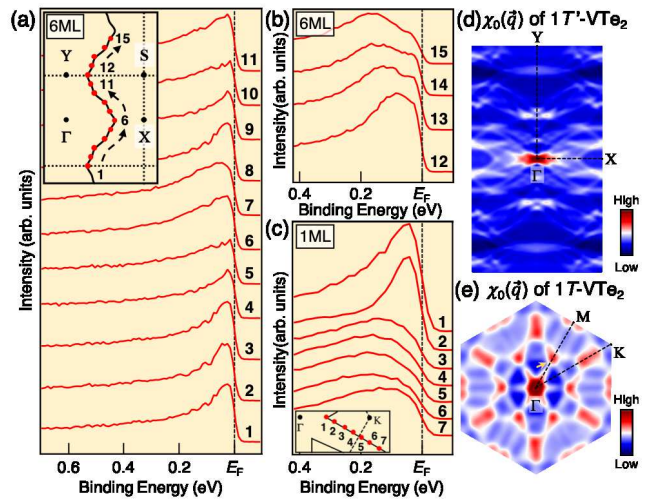}
\caption{(a), (b) EDCs near $E_{\rm{F}}$ at $T$ = 40 K for 6ML VTe$_2$ measured at various $k_{\rm{F}}$ points on the quasi-1D FS [points 1-11 and 12-15 (see inset), respectively]. (c) EDCs at the representative $k_{\rm{F}}$ points on the triangular FS for 1ML 1$T$-VTe$_2$ \cite{SugawaraPRBR2019}. (d), (e) Imaginary parts of electronic susceptibility for 1ML 1$T$'- and 1$T$-VTe$_2$, respectively, obtained from the calculated FSs in Figs. 2(c) and 2(e).}
\label{fig:fig4}
\end{center}
\end{figure}

Now we discuss the implication of present results in relation to the mechanism of structural transition. To examine a possibility that the 1$T$' phase in multilayer VTe$_2$ is triggered by the instability of FS toward the formation of 2$\times$1 CDW in the 1$T$ phase, we have calculated FS for a 2ML 1$T$-VTe$_2$ slab by taking into account that the 2ML film already takes the 1$T$' structure. The calculated FS, which consists of a triangular pocket at $\rm{K}$ and an elongated pocket at $\rm{M}$ (see Appendix \ref{App. G}), does not contain parallel segments spanned by the nesting vector with the 2$\times$1 periodicity, as also confirmed by the calculated electronic susceptibility. Moreover, judging from the absence of corresponding band folding (see Fig. 11 in Appendix \ref{App. H}), our ARPES data are also incompatible with the formation of 2$\times$1 CDW. These results do not support the FS-nesting-driven structural transition. Thus, the mechanism of structural transition in ultrathin VTe$_2$ would be different from that of bulk showing the 1$T$''-to-1$T$ phase transition at 482 K, because the bulk 1$T$'' phase can be associated with the formation of CDW \cite{MitsuishiNatureCommun2020}.

As an alternative scenario to explain the structural transition, the electron occupancy of the V 3$d$ band may play an important role \cite{CoelhoJPhysChemLett2019, SoulardJSolidStateChem2005, FengAPL2016}, because it is known to be sensitive to the crystal structure. It has been suggested from x-ray photoemission spectroscopy and STM studies on multilayer VTe$_2$ films that the charge transfer from Te to V atoms associated with the Te-Te interlayer coupling takes place upon increasing the number of layers from monolayer to bilayer \cite{CoelhoJPhysChemLett2019}. Our band-structure calculations also support the reduced (increased) occupancy of Te (V) orbital in 1$T$'-VTe$_2$, as seen from the clear $E_{\rm{F}}$-crossing of Te 5$p$ band in Fig. 2(f) and resultant emergence of a new hole pocket in Fig. 2(e), in contrast to the absence of Te-5$p$-derived FS in the calculations for the 1$T$ system [Figs. 2(d) and 2(c)]. On the other hand, our ARPES data suggest that, although there exists an energy shift of the Te 5$p$ band between 1 and 6ML [compare Figs. 1(e) and 1(f); the shift is better visualized in the EDCs at $\Gamma$ overlaid in the intensity plot], the Te 5$p$ band for 6ML does not apparently cross $E_{\rm{F}}$. This indicates that the Te 5$p$ orbital is fully occupied in both 1 and 6ML, implying that the simple Te-to-V charge-transfer scenario does not satisfactorily explain the occurrence of structural transition.

Although an accurate estimation of the actual V electron occupancy is difficult due to the spectral broadening and the complication from multiple domains, it is inferred that the V electron occupancy for 6ML is likely larger than that for 1ML, because the experimental quasi-1D FS [dashed curve in Fig. 3(a)] coincides with the calculated V-3$d$-derived quasi-1D FS shown in Fig. 2(e); according to the Luttinger's theorem, the total V-electron count that reflects the area of the calculated quasi-1D FS must be larger than the half filling since the Te-5$p$-derived large hole pocket exists in the calculation. It is tempting to consider a possible V intercalation in multilayer films, because charge transfer from the substrate was found to be negligible. It has been reported that V atoms can be intercalated into the epitaxial multilayer VTe$_2$ film and show the 2$\times$1 periodicity (corresponding to V$_3$Te$_4$) \cite{LasekACSNano2020}. By considering this point, we have calculated the bulk-band structure for V$_3$Te$_4$ (see Fig. 12 in Appendix \ref{App. I}) and found that the top of Te 5$p$ band is located far below $E_{\rm{F}}$ due to the excess electron doping from intercalated V atoms to VTe$_2$ layers. Obviously, our ARPES data do not support the formation of stoichiometric V$_3$Te$_4$ film because the top of Te 5$p$ band is almost at $E_{\rm{F}}$ in the experiment. This implies the partial intercalation of V atoms, although such atoms are not clearly resolved by our cross-sectional TEM measurements probably because it is difficult to visualize randomly distributed small amount of V atoms (see Appendix \ref{App. C}). Such intercalated V atoms can be responsible for the possible electron doping to the V 3$d$ band in the multilayer films and switch the free-energy balance between the 1$T$ and 1$T$' phases to lead to the structural transition. The present result implies that the accurate tuning of interlayer effect and orbital occupancy is important for controlling the crystal and electronic phases in ultrathin TMDs.

Finally, we discuss how the CDW evolves from the 4$\times$4 periodicity in 1ML VTe$_2$ to the 3$\times$1$\times$3 one in bulk VTe$_2$. Intriguingly, our multilayer films show the 2$\times$1 periodicity at least up to 14ML, distinct from the 3$\times$1$\times$3 periodicity appearing in the bulk. We speculate that there are two possibilities to account for this difference. First possibility is that, although the 3$\times$1$\times$3 CDW requires the 3D Fermi-surface-nesting condition associated with a finite wiggling of quasi-2D Fermi surface in the $k_{\rm{z}}$ direction, this condition is not satisfied in our film because the film is still thin enough to be regarded as 2D. Second possibility is an essential difference in the growth condition between MBE and bulk single-crystal growth. Partially intercalated V atoms (although they are difficult to resolve by TEM) may cause an electron doping to the V 3$d$ band and stabilize the 1$T$' phase. On the other hand, the situation of V intercalation may be different for bulk crystals and the condition to stabilize the 1$T$' phase may not be satisfied. These points need to be addressed by further experiments such as the partial intercalation to bulk VTe$_2$ crystals and fabrication of very thick films by MBE.

\section{CONCLUSION}
To summarize, we have reported an ARPES study of monolayer and multilayer 1$T$-VTe$_2$ films on bilayer graphene. We have revealed a drastic change in the FS topology and the band structure between the 1 and 6ML films, likely associated with the 1$T$-to-1$T$' structural transition. We also found the metallic state on entire quasi-1D FS at low temperature in 6ML film, in contrast to the 1ML counterpart characterized by pseudogap opening. The present result demonstrates an important role of interlayer effect for controlling the crystal phase, and paves a pathway torward understanding the interplay between the fermiology, structural phase, and electronic properties in ultrathin TMDs.

\begin{acknowledgments}
We thank K. Horiba and H. Kumigashira for their assistance in the ARPES experiments. This work was supported by JST-CREST (No. JPMJCR18T1), JST-PREST (No. JPMJPR20A8), JSPS KAKENHI Grants (Nos. JP18H01821, JP18H01160 and JP17H01139), KEK-PF (Proposal No. 2018S2-001), Science research projects from Shimazu Science Foundation, World Premier International Research Center, Advanced Institute for Materials Research. T. Kawakami and T. Kato acknowledge support from GP-Spin at Tohoku University.
\end{acknowledgments}

\appendix
\section{SAMPLE FABRICATION, ARPES MEASUREMENTS, AND BAND CALCULATIONS}\label{App. A}
Monolayer and multilayer VTe$_2$ films were grown on bilayer graphene by the molecular-beam-epitaxy (MBE) method. Bilayer graphene was fabricated by direct resistive heating of a $n$-type Si-rich 4$H$-SiC(0001) wafer at 1100 $^\circ$C at 1$\times$10$^{-9}$ Torr for 30 min. Then, VTe$_2$ film was grown by depositing V atoms on bilayer graphene in Te atmosphere. Number of layers are controlled by the deposition time while keeping the same deposition rate. The substrate temperature during the sample growth was kept at $\sim$ 300 $^\circ$C . The as-grown film was annealed at 300 $^\circ$C for 30 min and then transferred to the ARPES-measurement chamber directly connected to the MBE system. The growth process was monitored by the reflection high-energy electron diffraction (RHEED).  ARPES measurements were carried out using a Scienta-Omicron DA30 spectrometer with the beamline BL-28A at Photon Factory, KEK. Linier horizontal polarized light of 74 eV was used to excite photoelectrons. The energy and angular resolutions were set to be 20 meV and 0.2$^\circ$, respectively. The sample was kept at $T$ = 40 K during the ARPES measurements. The Fermi level ($E_{\rm{F}}$) of sample was referenced to that of a gold film deposited onto the sample substrate. A cross-sectional transition electron microscopy (JEM-ARM200F, JEOL) was operated at 200 kV. First-principles band-structure calculations for free-standing monolayer/multilayer 1$T$- and 1$T$'-VTe$_2$ were carried out by using the Quantum Espresso code package \cite{GiannozziJPhysCondesMatter2009} with generalized gradient approximation \cite{PerdewPRL1996}. Ultrasoft pseudopotential was used and the spin-orbit interaction was included in the calculations. The plane-wave cutoff energy and the $\textit{\textbf{k}}$-point mesh were set to be 70 Ry and 12$\times$12$\times$1, respectively. In the calculations, a vacuum layer with more than 15 $\rm{\AA}$ thickness was inserted between VTe$_2$ monolayers/multilayers. To evaluate the FS-nesting condition, the calculation of FS with the fine $\textit{\textbf{k}}$-point mesh was performed by using Wannier90 code \cite{PizziJPhysCondes2020}.

\section{CALCULATED VALANCE-BAND STRUCTURE FOR 1 AND 6ML 1$T$-VTe$_2$}\label{App. B}

\begin{figure}[htbp]
\begin{center}
\includegraphics[width=3.2in]{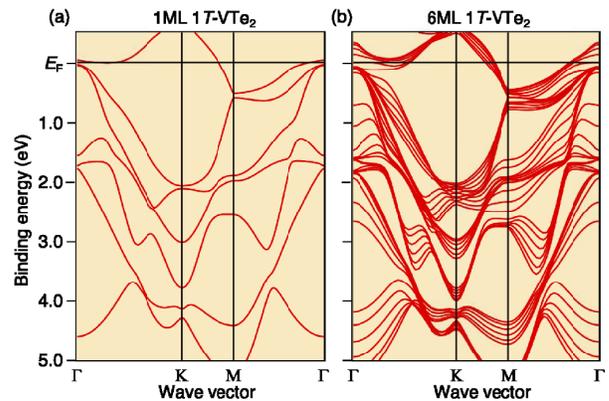}
\caption{Calculated band structure along high-symmetry cuts for (a) 1ML and (b) 6ML 1$T$-VTe$_2$.}
\label{fig:fig5}
\end{center}
\end{figure}

\begin{figure}[htbp]
\begin{center}
\includegraphics[width=3.2in]{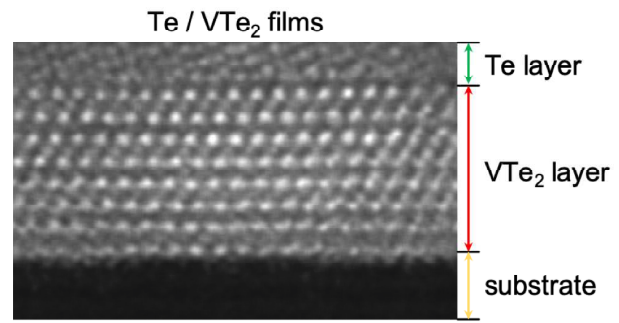}
\caption{Cross-sectional TEM image for multilayer (4ML) VTe$_2$ film. After the MBE growth, the VTe$_2$ film was capped by a few nm-thick Te layer to avoid the degradation of sample surface during transportation and TEM measurements. Note that the film thickness was 4ML but not 6ML because the deposition condition was slightly different from that for the ARPES measurement.}
\label{fig:fig6}
\end{center}
\end{figure}

We adopted the calculation for 1ML but not for 6ML to compare with the ARPES data, because the inclusion of multilayers in the slab calculations produces many bands whose individual dispersions cannot be easily resolved in the experiment. To discuss the overall agreement or disagreement between the experiment and calculation, it is still useful to compare with the calculation for 1ML. Here, to support this argument and to clarify the possible number-of-layer dependence of electronic states in 1$T$-VTe$_2$, we have performed first-principles band-structure calculations for free-standing 1 and 6ML slabs of 1$T$-VTe$_2$, and show the result in Fig. 5. Although band structure for 6ML is complicated by the presence of many bands, one can see overall agreement in the band structure between 1 and 6ML due to the inherently quasi-two-dimensional (2D) nature of the electronic states in 1$T$-VTe$_2$. We found that the experimental data for the 6ML film in Fig. 2(b) show several disagreements with the calculated band structure for 6ML 1$T$-VTe$_2$ in Fig. 5(b). This suggests that our 6ML film does not take a 1$T$ structure.

\section{CROSS-SECTIONAL TEM MEASUREMENT FOR MULTILAYER VTe$_2$ FILM}\label{App. C}

To clarify the crystal structure of multilayer 1$T$'-VTe$_2$ films, we performed a cross-sectional TEM measurement. As shown in Fig. 6, individual VTe$_2$ layer constituting of three monolayers; i.e. top and bottom layers from brighter Te atoms and middle layers from darker V atoms, can be clearly identified. As can be seen, each VTe$_2$ layer is separated from adjacent VTe$_2$ layers via van der Waals gap and no periodically ordered V atoms are found within the van der Waals gap, ruling out the formation of V$_3$Te$_4$ structure. It is noted that we were unable to distinguish the 1$T$ vs 1$T$' structure because the atomic displacement of the 1$T$' structure relative to the 1$T$ structure is not sufficiently large compared to the resolution of TEM.

\section{COMPARISON OF THE EXPERIMENTAL AND CALCLATED ENERGY CONTOURS}\label{App. D}
Figure 7 shows the ARPES intensity plots at several $E_{\rm{B}}$ slices ($E_{\rm{B}}$ = 0 to 1.4 eV by every 0.2 eV step) compared with the equi-energy contour of calculated bands for (a) 1ML 1$T$-VTe$_2$ and (b) 1ML 1$T$'-VTe$_2$. One can recognize that the matching between the ARPES intensity and the calculation for 1$T$ phase is poor, as highlighted by the obvious disagreement at higher $E_{\rm{B}}$ of 0.2-0.8 eV around the Brillouin-zone boundary, justifying that the 1$T$ phase is not realized in our 6ML film. As can be seen in Fig. 7(b), so many bands show up in the calculation for the 1$T$' phase, due to the doubling of unit cell and the existence of three crystal domains rotated by 120$^\circ$ from each other. Thus, it is not so easy to find a direct correspondence between the calculated equi-energy contours and ARPES intensity. Nevertheless, one may still find a reasonable agreement between them. For example, at $E_{\rm{B}}$ = 0.2 eV, the dark area with snowflake shape centered at the $\Gamma$ point of first Brillouin zone in the experiment appears to be correlated with the shape of calculated contours. Existence of bright intensity in the second Brillouin zones at ($k_x$, $k_y$) = ($\sim$0 $\rm{\AA}^{-1}$, $\sim$1 $\rm{\AA}^{-1}$) at $E_{\rm{B}}$ = 0.2-0.8 eV is consistent with the calculation that signifies some bands in that $\bf{k}$ region, distinct from the case of the calculation for the 1$T$ phase [Fig. 7(b)]. At the same time, some disagreements can be also found between the experiment and calculation. For example, at $E_{\rm{B}}$ = 1.2 eV, it is hard to attribute the bright area of ARPES intensity to the specific energy contour in the calculation. These results suggest that our multilayer films take the 1$T$' structure with three crystal domains rotated by 120$^\circ$ from each other.

\onecolumngrid
\begin{figure*}[htbp]
\begin{center}
\includegraphics[width=6.8in]{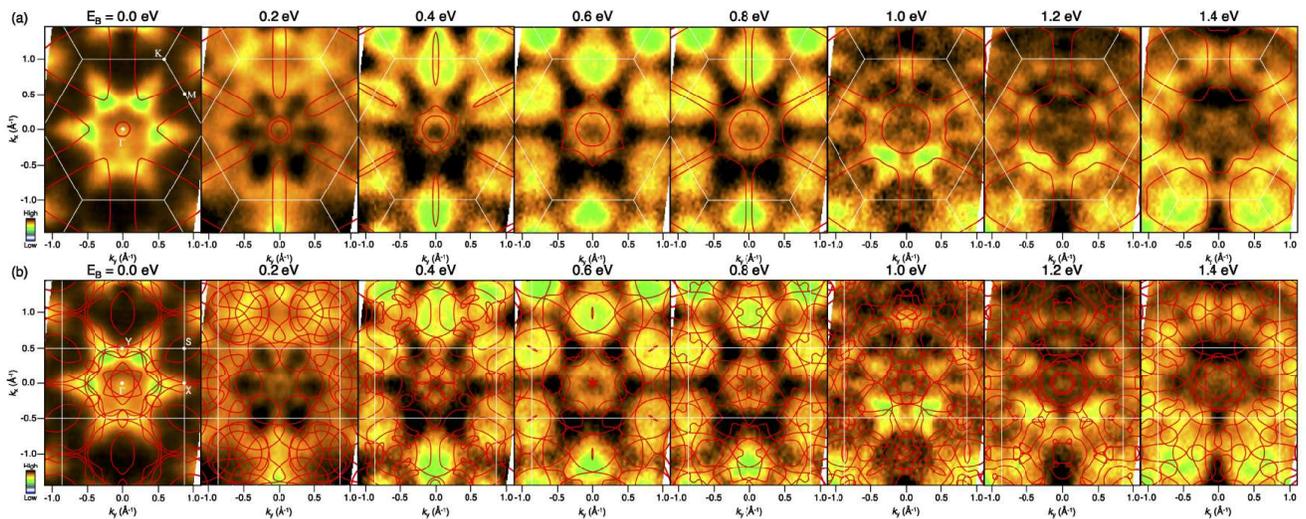}
\caption{ARPES intensity plots as a function of $k_x$ and $k_y$ obtained at several $E_{\rm{B}}$ slices for 6ML VTe$_2$ compared with the calculated equi-energy contours for (a) 1ML 1$T$-VTe$_2$ and (b) 1ML 1$T$'-VTe$_2$. The latter calculation takes into account three-crystal domains rotated by 120${}^\circ$ form each other.}
\label{fig:fig7}
\end{center}
\end{figure*}
\twocolumngrid

\onecolumngrid
\begin{figure*}[htbp]
\begin{center}
\includegraphics[width=5.6in]{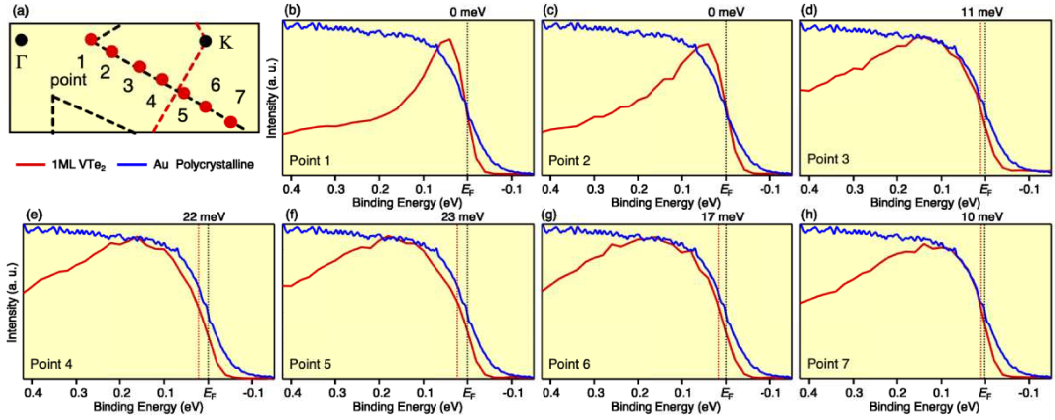}
\caption{(a) Schematic FS for 1ML 1$T$-VTe$_2$ together with the $\bm{k}$ points (red circles) at which EDCs were obtained. (b)-(h) EDCs at $T$ = 40 K at the representative \bm{k}$_{\rm{F}}$ points on the triangular FS compared to the EDC of polycrystalline Au film.}
\label{fig:fig8}
\end{center}
\end{figure*}
\twocolumngrid

\section{ANISTOROPIC PSEUDOGAP OF 1ML VTe$_2$}\label{App. E}

We compare in Fig. 8 the EDC at $T$ = 40 K for monolayer 1$T$-VTe$_2$ obtained at representative Fermi wave vector ($\bf{k}_{\rm{F}}$) points on the triangular Fermi surface centered at the K point, compared with that of polycrystalline Au film with a clear Fermi-edge cut-off. At points 1 and 2 at around the corner of the triangular pocket, one can recognize a sharp peak whose leading-edge midpoint well coincides with that of Au film, signifying the absence of energy gap (note that the steeper leading edge for VTe$_2$ is primarily associated with the sharp nature of the peak). On the other hand, at point 3, the leading edge of VTe$_2$ is slightly (by 11 meV) shifted toward higher binding energy with respect to that of Au due to the energy-gap opening. The leading-edge shift gradually becomes larger on approaching point 5 (a $\rm{\bf{k}}_F$ point along the KM line) and takes a maximum (23 meV) at this point, and then starts to decrease toward point 7. Such a systematic change supports anisotropic nature of the pseudogap.

\section{NUMBER OF LAYER DEPENDENCE OF RHEED IMAGES IN VTe$_2$  FILMS}\label{App. F}

Figure 9 shows the RHEED patterns obtained during the growth of VTe$_2$ films on bilayer graphene substrate. While the RHEED pattern of the 1ML film is characterized by the regular 1$\times$1 streaks originating from the 1$T$ structure (yellow arrow), one can see emergence of 2$\times$2 superstructure (white arrows) as soon as the film thickness reaches 1.5ML (note that ``1.5ML'' means the mixture of 1 and 2MLs). This suggests that the 1$T$' structure is already formed in the 2ML film. One can also see in Fig. 9 that the location of RHEED spot remains unchanged for the films thicker than 1.5ML. This suggests that the crystal always keeps the 1$T$' structure above 2ML.

\begin{figure}[htbp]
\begin{center}
\includegraphics[width=3.0in]{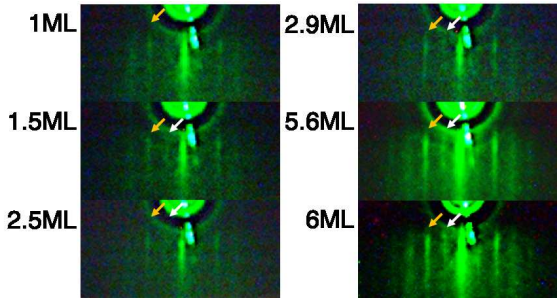}
\caption{Number-of-layer-dependent RHEED images of the VTe$_2$ films grown on bilayer graphene. The number of layers was estimated from the deposition time.}
\label{fig:fig9}
\end{center}
\end{figure}

\section{CALCULATED ELECTRONIC STATES AND ELECTRONIC SUSCEPTIBILITY FOR 2ML 1$T$-VTe$_2$}\label{App. G}

\begin{figure}[htbp]
\begin{center}
\includegraphics[width=3.0in]{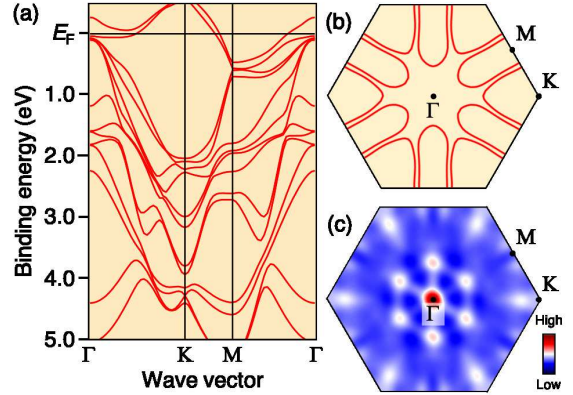}
\caption{(a), (b) Calculated band structure and FS, respectively, for 2ML 1$T$-VTe$_2$. (c) Intensity plot of the calculated imaginary part of electronic susceptibility $\chi_0(\bf{q})$ for 2ML 1$T$-VTe$_2$.}
\label{fig:fig10}
\end{center}
\end{figure}

To pin down the origin of 1$T$-to-1$T$' structural phase transition in VTe$_2$, we have performed the band calculations for 2ML 1$T$-VTe$_2$, and show the calculated band dispersion and Fermi surface (FS) in Figs. 10(a) and 10(b), respectively. Compared to the 1ML case [Fig. 1(c)], the calculated FS for 2ML is characterized by the existence of an additional elliptical pocket centered at the M point associated with the doubling of unit cell. Also, one can see that the volume of triangular FS at the K point is smaller than that of 1ML and thus the charge balance is kept unchanged. We found that the calculated FSs for 2ML are not well connected by the nesting vector with 2$\times$1 periodicity, as in the case of 1ML. This is also confirmed by the calculation of electronic susceptibility in Fig. 10(c) obtained from the calculated FS of 2ML that is characterized by the existence of sharp spots at $\sim$1/4 $\bf{G}$ ($\bf{G}$: reciprocal lattice vector) but not at $\sim$1/2 $\bf{G}$. These results suggest that the observed structural transition is unlikely driven by the instability of FS toward formation of 2$\times$1 CDW in 1$T$-VTe$_2$.

\section{COMPARISON OF ARPES INTENSITY BETWEEN 1ML AND 6ML FILMS}\label{App. H}

\begin{figure}[htbp]
\begin{center}
\includegraphics[width=3.5in]{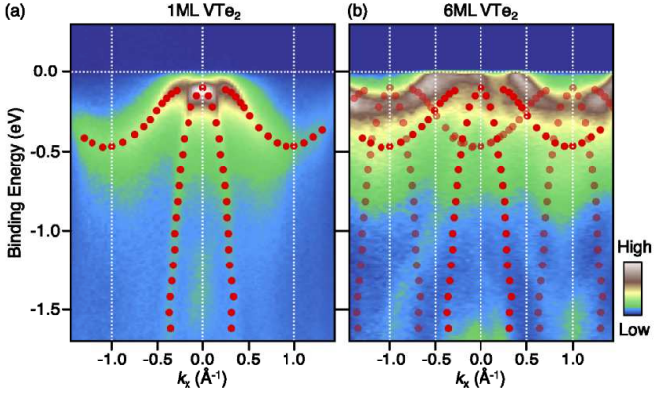}
\caption{(a), (b) ARPES intensity plots along the $k_x$ cut (the $\rm{\Gamma{M}}$ cut of 1$T$ Brillouin zone) measured at $T$ = 40 K for 1ML and 6ML films. Red dots in (a) show the experimental band dispersion for the Te 5$p$ and V 3$d$ bands obtained by tracing the peak position of EDCs. Experimental band dispersion in (a) is folded by assuming the 2$\times$1 Brillouin zone in (b).}
\label{fig:fig11}
\end{center}
\end{figure}

Figure 11 shows a comparison of the ARPES intensity along the $k_x$ cut (the $\rm{\Gamma{M}}$ cut of 1$T$ BZ) between the 1ML and 6ML films. If the 2$\times$1 CDW shows up in the 6ML film, the band dispersion of 1$T$ phase at the $\Gamma$ point is expected to be folded onto the M point of the 1$T$ Brillouin zone. As a result, one can expect the reconstructed band structure characterized by the replica bands of 1$T$-originated main bands as shown by red dots in Fig. 11(b). However, the matching between the expected replica bands and the ARPES intensity is obviously poor. For example, the V 3$d$ bands crossing $E_{\rm{F}}$ in the 6ML sample is located at much closer to $E_{\rm{F}}$ compared to the expected replica bands. Also, the dispersion of higher lying Te 5$p$ bands is not well reproduced by the replica bands. Such critical disagreements cannot be reconciled even after we consider the hybridization-gap opening at the band intersection associated with the 2$\times$1 periodic potential. This suggests that the observed 2$\times$1 periodicity in 6ML film cannot be interpreted in terms of the formation of 2$\times$1 CDW.
\vskip1\baselineskip
\vskip1\baselineskip

\section{BAND CALCULATIONS OF BULK V$_3$Te$_4$}\label{App. I}

\begin{figure}[htbp]
\begin{center}
\includegraphics[width=3.1in]{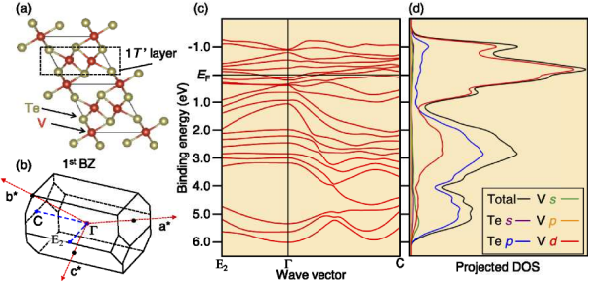}
\caption{(a), (b) Crystal structure and first Brillouin zone of of bulk V$_3$Te$_4$, respectively. Lattice parameters were obtained from material-project web site \cite{JainAPLmaterials2013}. (c), (d) Calculated band structure and corresponding partial density of states (DOS), respectively, for bulk V$_3$Te$_4$.}
\label{fig:fig12}
\end{center}
\end{figure}

We have performed the band calculations for V-intercalated VTe$_2$ with the 2$\times$1 V-atom order [corresponding to bulk V$_3$Te$_4$: see Fig 12(a) and Fig. 12(b) for the crystal structure and bulk Brillouin zone, respectively]. By looking at the calculated band structure in Fig. 12(c) together with the projected density of states in Fig. 12(d), we found that the V 3$d$ band is located within the binding-energy range of $\sim$ $\pm$1.2 eV, whereas top of the Te 5$p$ bands is at far below $E_{\rm{F}}$ ($E_{\rm{B}}$ $\sim$ 1 eV) due to the excess electron doping to the VTe$_2$ layers from the intercalated V atoms. Our ARPES data in Fig. 1(f) in the main text signify that top of the Te 5$p$ band is located almost at $E_{\rm{F}}$. This indicates that our epitaxial film does not form the stoichiometric V$_3$Te$_4$ film, although small amount of V atoms may be intercalated between the layers.

\end{document}